\documentclass[lettersize,journal]{IEEEtran}
\usepackage{cite}
\usepackage{amsmath,amssymb,amsfonts}
\usepackage{multirow}
\usepackage{amsthm}
\usepackage{mathrsfs}
\usepackage{xcolor}
\usepackage{textcomp}
\usepackage{manyfoot}
\usepackage{booktabs}
\usepackage{algorithm}
\usepackage{algorithmicx}
\usepackage{algpseudocode}
\usepackage{listings}
\usepackage{subcaption}
\usepackage{cleveref}
\usepackage{enumitem}
\usepackage{graphicx}
\usepackage{textcomp}
\def\BibTeX{{\rm B\kern-.05em{\sc i\kern-.025em b}\kern-.08em
    T\kern-.1667em\lower.7ex\hbox{E}\kern-.125emX}}

\begin{document}
%\history{Date of publication xxxx 00, 0000, date of current version xxxx 00, 0000.}
%\doi{}

\title{Graph neural network for in-network placement of real-time metaverse tasks in next-generation network}

\author{Sulaiman Muhammad Rashid, Ibrahim Aliyu, \IEEEmembership{Member,~IEEE}, IL-KWON Jeong, Tai-Won Um, and Jinsul Kim, \IEEEmembership{Member,~IEEE}
        % <-this % stops a space
\thanks{This work was supported by Innovative Human Resource Development for Local Intellectualization program through the Institute of Information \& Communications Technology Planning \& Evaluation(IITP) grant funded by the Korea government(MSIT). (IITP-2024-00156287) and this work was supported by the Institute of Information \& communications Technology Planning \& Evaluation(IITP) grant funded by the Korea government(MSIT) (RS-2024-00345030).\textit{Corresponding authors: Tai-Won Um (email: stwum@jnu.ac.kr) and Jinsul Kim (email: jsworld@jnu.ac.kr)}.}% <-this % stops a space
\thanks{Sulaiman Muhammad Rashid, Ibrahim Aliyu, and Jinsul Kim are with the Department of Intelligent Electronics and Computer Engineering, Chonnam National University, Gwangju 61186, Korea.}
\thanks{IL-KWON Jeong is with the Hyper-Reality Metaverse Research Laboratory Content Research Division, Electronics and Telecommunications Research Institute(ETRI), Daejeon, Korea}
\thanks{Tai-Won Um is with the Graduate School of Data Science, Chonnam National University, Gwangju 61186, Korea.}}

\maketitle
\IEEEpubid{\begin{minipage}{\textwidth}\ \\[30pt] % adjust space to prevent overlap
This work has been submitted to the IEEE for possible publication. Copyright may be transferred without notice, after which this version may no longer be accessible.
\end{minipage}}
\maketitle

\begin{abstract}

Delivering realistic and real-time virtual experiences in the Metaverse requires computationally intensive rendering tasks with stringent delay constraints. These demands pose significant challenges due to the dynamic nature of user requests and the heterogeneity of network resources. While traditional machine learning (ML) techniques have been applied to address task placement, they often fall short in dynamic and heterogeneous environments, failing to accurately model complex task-node relationships. This study proposes a GNN-based adaptive framework for in-network placement of delay-constrained tasks in the Metaverse, leveraging the Computing in the Network (COIN) paradigm. The task placement problem is formulated as an Integer Linear Programming (ILP) model to minimize system costs while satisfying strict delay requirements. Optimal solutions are initially obtained using standard optimization solvers; however, these solutions are computationally expensive and unsuitable for real-time applications. To address this, the solutions are used to train the GNN offline, enabling rapid and efficient placement decisions during runtime. By modeling both connected and isolated node relationships, the GNN framework dynamically adapts to network variability, achieving a placement accuracy of 92\%, compared to 76\% for multilayer perceptron (MLP) and 70\% for decision trees (DTs). Finally, the simulation results shows that the proposed model can achieve significant performance gains compared to other machine learning models, and that task splitting across multiple COIN nodes enables better utilization of network resources.

\end{abstract}


%\titlepgskip=-15pt

\section{Introduction}
\label{sec1}
%% Labels are used to cross-reference an item using \ref command.
The Metaverse represents a hypothetical “parallel virtual world,” embodying lifestyles and work environments within virtual cities as an alternative to future smart cities \cite{allam2022}. It envisions seamless integration into our daily lives, offering users immersive real-time experiences globally \cite{chengoden2023,ramadan2023,buhalis2023}. However, realizing this vision presents a substantial challenge in ensuring the metaverse operates seamlessly in real-time, irrespective of geographical distances \cite{xu2023}. To achieve high levels of immersion and realism, metaverse applications must render graphics, animations, and simulations in real time \cite{cheng2023} which is crucial for virtual reality (VR) and augmented reality (AR) applications aiming to induce a sense of presence and “being there” \cite{dhelim2022}. In the metaverse, users can generate and modify 3D digital content, where rendering plays a pivotal role in producing such content \cite{liu2022}, involving the computation of images perceived by users based on computer graphics principles \cite{https://doi.org/10.1111/cgf.14507}. Therefore, efficient handling of rendering tasks for real-time metaverse applications is crucial and imperative for research and development.

While mobile edge computing (MEC) offers a remedy through remote task offloading (TO), it struggles to meet extensive user concurrent demands \cite{10366259,9712216,ISLAM2021102225,Chen2023}. The computing in the network (COIN) paradigm emerges as a promising solution, utilizing untapped network resources to execute tasks, diminishing latency, and fulfilling quality of experience (QoE) requirements \cite{10366259,9606828}. Nonetheless, augmenting computing resources or enabling COIN leads to heightened power consumption. Effectively allocating COIN computing resources in real-time to adapt to continually shifting user demands while ensuring overall system availability poses a critical challenge.

\subsection{Motivation and Contributions}
Despite substantial advancements in handling time-intensive tasks, challenges persist, including potential bottlenecks arising from resource constraints and network congestion \cite{shakarami2020}. Previous researchers propose the concept of in-network computing (INC) as a solution for managing delay-constrained tasks \cite{kianpisheh2023,mai2021,lia2021}. This leverages the computing capability of nodes deployed across the INC paradigm. Deciding which INC node should execute a task or whether to offload it to the MEC due to network load can be achieved by adopting a software-defined networking (SDN) architecture \cite{priyadarsini2021}. SDN provides programmability, flexibility, and centralized control\cite{xie2015}. 

Within metaverse applications, rendering high-quality 3D content is computationally intensive \cite{lin2023}. It involves input processing, environment rendering, spatial mapping and tracking, and composition and display \cite{huzaifa2021,https://doi.org/10.1111/cgf.14507}. 3D rendering in the metaverse starts with capturing users and objects in 2D using RGBD sensors, e.t.c \cite{10.3389/frsip.2023.1139897}. Preprocessing steps such as environment rendering, spatial mapping, and tracking are performed at nodes near each user to minimize latency. The final rendering aggregates all preprocessed inputs into a unified 3D format, making it most efficient at a single node. Centralizing this step ensures consistency and synchronization for all participants. Therefore, initial preprocessing is treated as an independent task for distributed nodes, while the final rendering is executed at a single node for optimal performance.

However, achieving optimal decisions requires complex mathematical models, posing a challenge to efficient task allocation. As outlined in \cite{lia2022}, machine learning (ML) approaches can be devised to determine how to allocate computing tasks across the network. Current machine learning (ML) approaches exhibit low accuracy and struggle to model the complexities of INC environments effectively

Recently, graph neural networks (GNNs) have emerged as powerful tools for analyzing graph-structured data \cite{zhou2020,liang2022,gupta2021,isah2024graph}. Leveraging their capability, we designed a GNN-based model to address the dynamicity of task-node relationships in rendering tasks. Inspired by \cite{kipf2016semi}, we introduce a GNN-based Adaptive Network, a framework that models both independent and connected graph structures with two types of convolution operations: neighbor-aware and independent convolutions. Specifically, the neighbor-aware takes all the features from connected nodes to capture resource sharing relationships. However, isolated nodes may still be missed during classification. The independent convolution ensures that these isolated nodes still contributes to the task placement decisions through feature transformation.

%Early studies mainly focused on optimizing rendering algorithms to address the challenges of real-time rendering in metaverse applications, while other techniques, such as INC, which are drawing more researchers’ attention in delay-constrained computing, are yet to be devised for such problems. Previous works on INC focused on offloading tasks independently on SDN-based edge networks; none considered computing tasks on more than one node due to the complexity of being computed by a single node. Finally, different ML techniques have been utilized by various researchers to address dynamic task placement problems in SDN edge environments. However, GNN algorithms have not been applied to this field yet. Therefore, this study establishes a novel approach for studying placement and offloading decisions for delay-constrained tasks for real-time metaverse applications. This approach promises to employ and compare GNN with other ML techniques to efficiently distribute rendering tasks within an INC environment and impact the real-time rendering performance of metaverse applications.

Our work is applied to a metaverse concert scenario, where players interact and request rendering tasks distributed across heterogeneous INC nodes. The task placement problem is formulated as an Integer Linear Programming (ILP) model to minimize network resource usage while ensuring tasks meet strict execution deadlines. The ILP solutions are used to train the GNN model offline to enable fast, near-optimal task placement decisions during runtime.

\begin{itemize}
  %\item We propose an SDN-based network architecture that includes the control and data planes (comprising INC nodes, edge servers, and ingressors). The control plane receives requests from clients/group of players and orchestrates communication with the selected INC node/edge server for task execution.
  \item We formulate the optimal task placement problem through an Integer Linear Programming (ILP) to minimize rendering latency under task time execution constraints, queuing delays, and the limited computing capabilities of the executing nodes.
  \item We designed and implemented a GNN-based adaptive network model to identify the near-optimal placement of metaverse rendering tasks in real-time computing nodes, minimizing resource use while meeting delay requirements. 
  \item We conduct extensive experiments to evaluate the performance of the model results against some benchmark solutions. The performance of our proposed model surpasses other schemes in the context of delay-constrained computing in metaverse applications.
\end{itemize}

The rest of the paper is organized as follows: Section \ref{section2} addresses related studies, Section \ref{section3} presents the system model and Section \ref{section4} is our optimization problem. In Section \ref{section5}, we present optimal task placement using GNN; and Section \ref{section6} presents the experiments and evaluations to validate the results of our proposed study. Section \ref{section7} concludes the paper.

\section{Related work}\label{section2}

Recent studies have addressed task placement in edge computing, exploring both single-user systems, where tasks are either processed locally or offloaded to edge servers \cite{7914660}, and multi-user systems, where a single edge server handles multiple clients \cite{8847369,7307234}. In next-generation networks, task placement becomes more complex as real-time, compute-intensive tasks may need to be distributed across multiple edge servers. Challenges include managing distributed edge and centralized cloud systems, identifying optimal task executors, and allocating resources efficiently in dynamic environments, particularly with mobile users \cite{8847369,9113305}.

In task placement problems, key metrics such as latency, energy consumption, and bandwidth usage are often considered, either individually or collectively \cite{7906521,8674548}. Beyond dedicated servers, tasks can also be executed at network edge nodes enhanced with computing capabilities\cite{10366259,zheng2023dinc}. This concept of treating network nodes as computing resources is central to the in-network computing (INC) paradigm. In such environments, the presence of numerous candidate executors with varied and limited computing capabilities makes task placement decisions highly complex. Poor decisions can fail to meet user requirements and reduce system efficiency. As such, we categorize the related work into two parts as follows:

\subsection{In-network placement of delay-constraint tasks}

The future of computing is poised at the edge \cite{9052677,9083958}, where data-driven decisions necessitate near-instantaneous processing. Extensive research has delved into the task placement problem in edge computing environments. \cite{9171497,electronics8101076}. While edge computing offers a remedy through remote task offloading, it struggles to meet extensive concurrent user demands \cite{10366259,9712216}. Recent research focuses on in-network computing \cite{lia2021, 8487419,9495935}, where untapped network resources are utilized to execute tasks, diminishing latency and improving the quality of experience. However, in the literature, managing the limited computing resources is overlooked as tasks are fully offloaded to the nodes, leading to more delays. These methods have shortcomings for applications with high computational demand, like the metaverse application. Our work addresses these limitations by considering the splitting of tasks between two INC nodes or offloading them entirely to Multi-access edge computing server (MEC) to ensure that each tasks are executed within their time constraint.

\subsection{ML for SDN-based task placement}

The separation of the control and data planes facilitated by SDN has introduced a dynamic technique that enhances network flexibility  \cite{SDNNitheesh}. SDN-based task placement has been addressed by \cite{lia2021} through standard optimization solvers, however it requires more time to arrive at feasible solution.

In recent studies, ML approaches for task placement in SDN computing are gaining traction \cite{lia2022, shakarami2022, 9493245}. For instance, in \cite{shakarami2022}, authors proposed a deep learning-based resource allocation for SDN-enabled Fog architecture. Another work explored using supervised learning techniques in an intelligent edge for SDN-based task placement \cite{lia2022}. However, these approaches often fail to deliver high accuracy, particularly in complex, heterogeneous environments. Graph Neural Networks (GNNs) have recently shown promise in addressing these challenges by modeling graph-structured data \cite{9796910,9289378}. Their ability to generalize across diverse topologies and adapt to dynamic network conditions makes them suitable for task placement problems. However, current GNN models often overlook the contribution of isolated nodes, leading to suboptimal placement decisions. This limitation, combined with the inherent complexity of INC environments, hinders existing ML and GNN models from achieving the accuracy required for delay-constrained applications like real-time metaverse rendering. 

Our study addresses these gaps by leveraging a refined graph convolutional network (GCN) approach tailored for dynamic task placement in metaverse environments. By modeling both connected and isolated node relationships, we improve the accuracy and adaptability of task placement decisions, enabling efficient execution of delay-sensitive tasks even in highly variable and heterogeneous networks.

%Even though GNN requires longer training time, it can generalize across different network topologies and dynamic network conditions. This is why, in our analysis, we focus on the GNN model, in particular, the Graph Convolution Network (GCN) approach. Unlike the related literature, we considered a more challenging scenario of real-time metaverse application edge domain
%with multiple heterogeneous INC nodes possessing computing capabilities.

\section{System Model}\label{section3}
\label{sec3}
\begin{figure}[h]
  \centering
  \includegraphics[width=1.0\linewidth]{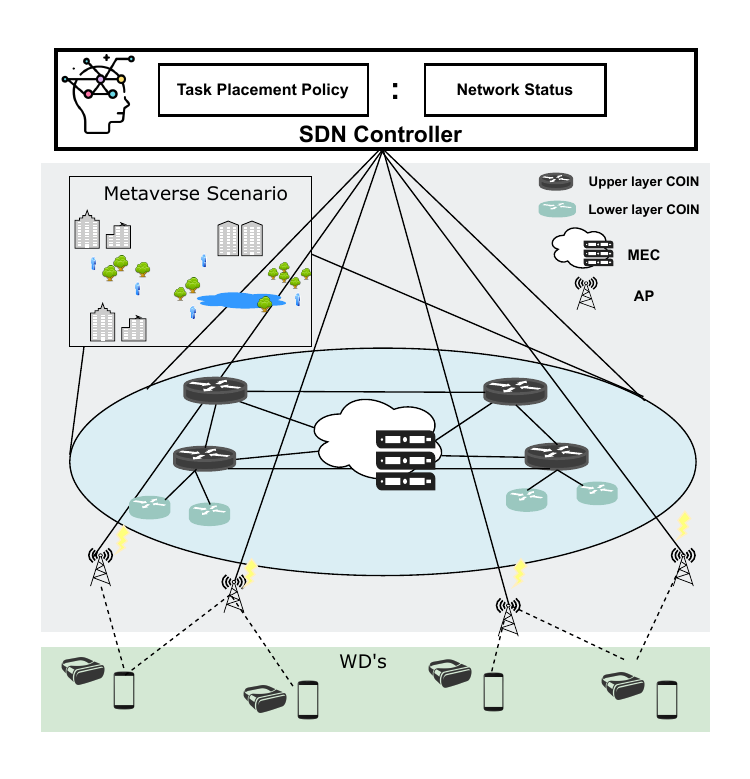}
  \caption{COIN-MEC metaverse task offloading scenario}
  \label{fig1:Experimental_Setup}
\end{figure}

In our scenario, (see Fig.\ref{fig1:Experimental_Setup}), set of WDs $\mathcal{N}= \{1,2,3 \dots N\}$  request computationally intensive tasks through $\mathcal{A}= \{1,2,3 \dots A\}$ set of Access points (APs). WDs can offload their tasks to a set $\mathcal{C}= \{1,2,3 \dots C\}$ of COIN or to a set $\mathcal{E}= \{1,2,3 \dots E\}$ of MEC. COIN and MEC form the set $\mathcal{M} \triangleq \mathcal{C} \cup \mathcal{E}$ of computing edge nodes. The SDN controller initializes resource information for $\mathcal{M} = \{1,2,3,...,M\}$, with computing capability \(F_i\) expressed in CPU cycles per second. Upon receiving a rendering request, the SDN controller returns the address \(A_k\) of selected nodes and establishes a communication link with nodes responsible for task execution. 

\subsection{Task Model}
  The WDs request $\mathcal{K}= \{1,2,3 \dots K\}$ set of rendering task, characterized by $(S_k, P_k, D^{\text{max}}_{k})$, where $S_k$ is the size of task $k$ in megabytes, $P_k$ denotes the workload (processing power) required in CPU cycles to accomplish the task, and $D^{\text{max}}_{k}$ signifies the maximum delay constraint of the rendering task. Unlike previous works \cite{8254044,8274943} focusing on atomic tasks, we considered metaverse divisible tasks. For simplicity, we assume tasks can be divided into independent joint subtasks. Therefore, for WD $i$, a task of type $k$ can either be Partial Offload (PO), i.e divided and offloaded to $c_{0}^k$ and $c_{1}^k$ corresponding to first and second COIN, or Full Offload (FO) i.e offloaded as a whole to MEC $e^k$. We therefore define the set of feasible decisions for WD $i$ as $\mathcal{D} \triangleq \{d_{PO}\} \cup \{d_{FO}\} $ and we use $d_i \in \mathcal{D}$ to indicate the decision for WD $i$'s task (i.e, $d_i = d_{PO}$ is the decision for partial offloading, and $d_i = d_{FO}$ is the decision for full offloading.)

The total time $T_i$ required by node $i$ to complete the task $k$ in $c_{0}^k$ and $c_{1}^k$ is given by: 

\begin{equation}
T^{c}_i = \frac{P_k}{F_{i,0}^c + F_{i,1}^c}, \quad i \in \mathcal{E}
\label{eq:equation1}
\end{equation}

while in $e^k$ is given as:

\begin{equation}
T^{e}_i = \frac{P_k}{F_{i}^e}, \quad i \in \mathcal{E}
\label{eq:equation2}
\end{equation}

where ${F_{i,0}^c \text{and} F_{i,1}^c}$ are the computing capabilities of the $c_{0}^k$ and $c_{1}^k$ respectively, and ${F_{i}^e}$ is the computing capability of $e^k$

Consistent with previous studies \cite{priyadarsini2021,lia2022}, the arrival rate of requests for rendering task $k$ at node $i$ follows a Poisson distribution characterized by $\alpha_k$ and can be defined as:

\begin{equation}
\sum_{k=\mathcal{K}}X_{i,k}\alpha_k
\label{eq:equation2}
\end{equation}

Where $X_{i,k}$ is the binary decision variable. It is equal to 1 if the task $k$ is executed by edge node $i$ and equal to 0 otherwise.

The allocation of computing resources received per task has an average distribution denoted by $\bar{F}$; Consequently, all edge nodes can collectively form an M/M/1 queuing model to process rendering tasks \cite{vijayashree2018}. The total computation delay for task $k$ at $c_{0}^k$ and $c_{1}^k$ is given by equation \ref{eq:equation3} while at $e^k$ is given by equation \ref{eq:equation4}:

\begin{equation}
\bar{D}_{i}^c = \frac{1}{\left( \frac{F_{i,0}^c + F_{i,1}^c}{\bar{F}} - \sum\limits_{k \in \mathcal{K}} X_{ik} \alpha_k \right)}, \quad i \in \mathcal{M}
\label{eq:equation3}
\end{equation}

\begin{equation}
\bar{D}_{i}^e = \frac{1}{{\left(\frac{{F_{i}^e}}{\bar{F}} - \sum\limits_{k=\mathcal{K}}X_{ik}\alpha_k \right)}}, \quad i \in \mathcal{M}
\label{eq:equation4}
\end{equation}

\subsection{Computing Model}

For notational convenience let us define the indicator function for WD $i$:

\begin{equation}
{I}(d_{PO}, d_{FO}) = \begin{cases}
1 & \text{if } d_{PO} = d \\
0 & \text{if } d_{FO} = d
\end{cases}
\end{equation}

Where $d_{PO}$ and $d_{FO}$ indicates decisions for \textbf{PO} and \textbf{FO} of task $k$ respectively.

\textbf{Network Cost:} For \textbf{PO}, the total network cost $NC$ is determined by the distance $h_{c}^0$ between the AP and $c_{0}^k$, the distance $h_{c}^1$ between $c_{0}^k$ and $c_{1}^k$, and size $S_k$ of the task $k$. While in the case of \textbf{FO}, distance $h_e$ between the AP and $e^k$ and the size $S_k$ of task $k$ is considered.

\begin{equation}
C^n = \alpha_k \sum_{k=1}^{K} S_k ((h_{c}^{0} + h_{c}^{1}) + h_{e}) I(d_{PO},  d_{FO}))
\label{eq:equation7}
\end{equation}

The number of exchanged nodes substantially impacts the network cost, potentially resulting in high traffic within the network. Finally, the system cost is expressed as:

\begin{equation}
C = \sum_{i\in {\mathcal{E}}} \sum_{k\in K} X_{ik} \cdot C^n
\end{equation}

\begin{table}[h]
    \centering
    \caption{Symbol Descriptions}
    \label{tab:symbols}
    \resizebox{\columnwidth}{!}{%
    \begin{tabular}{cl}
        \hline
        \textbf{Symbol} & \textbf{Description} \\
        \hline
        \(\mathcal{K}\) & Set of rendering tasks \\
        \(\mathcal{C}\) & Set of COINs \\
        \(F_i\) & Computing capacity of node \(i\) \\
        \(S_k\) & Size of task \(k\) (MB) \\
        \(P_k\) & Required processing power for task \(k\) \\
        \(D^{\text{max}}_{k}\) & Maximum delay constraint for task \(k\) \\
        \(T_{i}^c\) & Total time to execute task \(k\) in COIN \\
        \(T_{i}^e\) & Total time to execute task \(k\) in edge cloud \\
        \(\bar{F}\) & Average computation time \\
        \(h_{c}^{0}\) & Distance between AP $a$ and $c_{k}^0$ \\
        \(h_{c}^{1}\) & Distance between AP $c_{k}^0$  and $c_{k}^1$ \\
        \(h_{e}\) & Distance between AP $a$ and $e_{k}$ \\
        \(\alpha_k\) & Arrival rate of task \(k\) \\
        \hline
    \end{tabular}}
\end{table}

\section{Optimization Problem Formulation}\label{section4}
Our goal is to minimize the system cost by finding the optimal placement decision of task $k$. From the above formulation, the problem can be expressed mathematically as an integer linear programming (ILP) problem and is formulated as:

\begin{align}
\mathcal{P:} & \quad \min \sum_{i\in {\mathcal{E}}} \sum_{k\in K} X_{i,k} \cdot C^n \label{eq:P} 
\end{align}

s.t. :
\begin{align}
\text{} & \quad \sum_{d \in \mathcal{D}_i} {I}(d_{PO}, d_{FO})=1, \quad  \tag{9a} \label{eq:C1} \\
\text{} & \quad X_{i,k} \bar{D}_{i}^c \leq X_{i,k} \bar{D}_{i}^e \tag{9b} \label{eq:C2} \\
\text{} & \quad X_{i,k} \bar{D}_{i}^c \leq X_{i,k} \bar{D}_{i}^{max}  \tag{9c} \label{eq:C3} \\
\text{} & \quad {(F_{i,0}^c + F_{i,1}^c)} / {\bar{F}} - \sum\limits_{k \in \mathcal{K}} X_{ik} \alpha_k > 0  \tag{9d} \label{eq:C4} \\
\text{} & \quad X_{ij} \in \{0, 1\}, \quad i \in \mathcal{M}, \quad k \in \mathcal{K}  \tag{9e}
\label{eq:C5}
\end{align}

Constraint (\ref{eq:C1}) ensures that each WD task is partially offloaded to two COINS $c_{0}^k$ and $c_{1}^k$ or fully offloaded to the edge cloud $e^k$. Constraint (\ref{eq:C2}) ensure that the total computation time for \textbf{FO} is not greater than the computation time for \textbf{PO}. Equations (\ref{eq:C3}) confirm that the computing resources allocated to a task are within limits. The constraint in Equation (\ref{eq:C4}) ensures that the average service rate of edge nodes to be greater than the average task arrival rate in the case of offloading \textbf{PO}. Table \ref{tab:parameter_settings} presents the simulation parameter settings except otherwise stated.

\begin{figure}[t]
  \centering
  \includegraphics[width=1.0\linewidth]{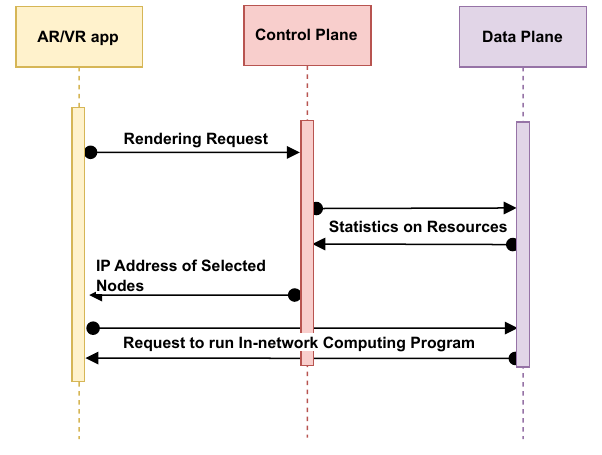}
  \caption{Sequence diagram of rendering request and INC procedures}
  \label{fig:sequence_diagram}
\end{figure}

\section{Optimal task placement using GNN}\label{section5}
The ILP problem in \cref{eq:P} corresponds to the Generalized Assignment Problem (GAP) \cite{CATTRYSSE1992260,lia2022}, which is NP-hard problem.  In practical scenarios, solving this problem using standard optimization solvers involves exhaustive searches to determine the optimal task placement, rendering such approaches computationally infeasible. Consequently, a more dynamic and efficient solution is required to address the ILP problem, even if this comes at the cost of achieving a near-optimal solution. To this end, we reformulate the task placement problem as a multi-class classification task.

In this study, we propose using a graph neural network (GNN) algorithm trained on labeled data derived from optimal decisions obtained using a standard optimization solver. The GNN model aims to predict the selected executor node corresponding to a given task request $k$. This model is deployed as a network application within the Software-Defined Networking (SDN) controller, which records various features that can be used as input for the classification.

\begin{table}[ht]
    \centering
    \caption{Parameter Settings}
    \label{tab:parameter_settings}
    \begin{tabular}{p{3.5cm}|p{4.0cm}}
        \hline
        \textbf{Parameter} & \textbf{Setting} \\
        \hline
        Computing capability ($F_i$) & 
        \begin{tabular}[c]{@{}l@{}}
            Lower layer COINs: $5 \times 10^8$ CPU \\cycles/s\\
            Upper layer COINs: $1 \times 10^9$ CPU \\cycles/s\\
            MEC: $1 \times 10^{10}$ CPU cycles/s
        \end{tabular} \\
        Task Size ($S_k$) & $10$ MB \\
        Average processing power required ($P_k$) & $1 \times 10^7$ CPU cycles\\
        Maximum delay constraint ($D_{k}^{\text{max}}$) & Uniformly distributed [10, 150] ms \\
        Arrival rate of task ($\alpha_k$) & $10$ requests/s \\
        \hline
    \end{tabular}
\end{table}

We define the \textbf{inputs} to the model using the task request vector $\overline{V_k}$ for a network with $\mathcal{M}$ potential executor nodes. Each vector is characterized by $F$ features, denoted as $\overline{V_k} = (v_{k,1}, v_{k,2}, \ldots, v_{k,F})$. The number of features is determined as $F = 2 \cdot |C^n| + 1$, where:

\begin{itemize}
    \item $v_{k,1}$ represents the maximum computation delay of the task $v_k$.
    \item $v_{k,n+1}$, where $n = [1, \ldots, |C^n|]$, represents the cost of executing the $k$-th task on node $i$, as defined in \cref{eq:equation3}.
    \item $v_{k,n+|C^n|+1}$, where $n = [1, \ldots, |C^n|]$, is a boolean vector with $|C^n|$ elements that encodes the constraint expressed in \cref{eq:C4}.
\end{itemize}

For each task execution, the algorithm produces an \textbf{output} specifying the selected node(s) as the executor(s) for task $k$. Once the executor is identified, the SDN controller communicates the address of the selected node(s) to the WD to facilitate task assignment.

\subsection{Graph Construction}

To formalize this problem, we give the mathematical symbol \(\mathcal{G}=(\mathcal{V},\mathcal{E})\) of the graph, where $\mathcal{V}$ is the set of nodes and $\mathcal{E}$ is the set of edges. The edges on $\mathcal{E}$ are constructed based on shared resources, i.e task $i$ and $j$ share an edge if they are located at the same node. The graph structure is encoded as an adjacency matrix $\mathcal{A} \in \mathbb{R}^{N x N}$, where:

\begin{equation}
 \mathcal{A}_{i,j} = \begin{cases}
1 & \text{if task $i$ and $j$ share same node}  \\
0 & \text{Otherwise } 
\end{cases}   
\end{equation}

Each node \(i \in \mathcal{V}\) is associated with a feature vector 
\({\psi}_i^{(0)} \in \mathbb{R}^F\), where \(F\) consist of all the input features. Our goal is to classify each task \(k\) into one of the classes, corresponding to the optimal offloading decision. 

\subsection{Feature Embedding}

The node feature vector \( \psi_i^{(0)} \) is embedded into a dense latent representation using a learnable embedding layer expressed as:

\begin{equation}
\psi_i^{(1)} = \sigma(W \cdot \psi_i^{(0)} + b)
\end{equation}

where \( W \in \mathbb{R}^{F \times \delta} \) is the weight matrix, \( b \in \mathbb{R}^{\delta} \) is the bias vector, \( \sigma(\cdot) \) is the activation function (ReLU), and \( \delta \) is the dimension of the embedding space.

\subsection{Graph Convolution}
The embeddings \( \psi_i^{(1)} \) are refined based on the structure of the graph. Two cases arise based on whether tasks \( i \) and \( j \) share resources.

\subsubsection{Neighbor-Aware Convolution}
When $\mathcal{A}_{i,j} = 1$ (i.e., they are connected in the graph), the convolution operation aggregates features from neighboring nodes and is defined as:
\begin{equation}
   \psi_i^{(2)} = \sigma\left(\sum_{j \in N(i)} \frac{1}{\sqrt{\deg(i) \cdot \deg(j)}} W^{(2)} \psi_j^{(1)} \right) 
\end{equation}

where:
    \( N(i) \) represents the set of neighbors of node \( i \),
    \( \deg(i) \) denotes the degree of node \( i \) (number of connected edges),
    \( W^{(2)} \) is the weight matrix for the graph convolutional layer,
    \( \sigma(\cdot) \) is the activation function (ReLU).

The adjacency matrix \( \mathcal{A} \) is normalized to account for node degrees using the formula:
\begin{equation}
    \tilde{\mathcal{A}} = D^{-1/2} \mathcal{A} D^{-1/2}
\end{equation}

where \( D \) is the degree matrix of \( \mathcal{A} \).

\subsubsection{Independent Convolution}
When $\mathcal{A}_{i,j} = 0$ (i.e., the graph is disconnected), the convolution operates independently for each node. In this case, the update is given by:
\begin{equation}
    \psi_i^{(2)} = \sigma(W^{(2)} \psi_i^{(1)})
\end{equation}

This ensures that even isolated nodes undergo feature transformation and contribute to the final classification \cite{kipf2016semi}.

\subsection{Output Layer for Classification}

The final node embeddings \( \psi_i^{(L)} \) after \( L \) GCN layers are passed through a softmax layer to classify each task \( i \) into one of $d_i \in \mathcal{D}$ offloading decisions (the selected COINs or MEC) expressed as:

\begin{equation}
p_i = \text{softmax}(W^{(L)} \psi_i^{(L)} + b^{(L)})
\end{equation}

where \( W^{(L)} \in \mathbb{R}^{\delta \times \mathcal{D}} \) and \( b^{(L)} \in \mathbb{R}^{\mathcal{D}} \) are the weights and biases for the output layer.

\subsection{Objective Function}

The training process minimizes the cross-entropy loss for all nodes:

\begin{equation}
\mathcal{L} = -\frac{1}{N} \sum_{i=1}^N \sum_{c=1}^C y_{i,c} \log(p_{i,c}),
\end{equation}

where \( y_{i,c} \) is the true label for node \( i \) and class \( c \), and \( p_{i,c} \) is the predicted probability for the same.

\begin{figure}[t]
  \centering
  \includegraphics[width=1.0\linewidth]{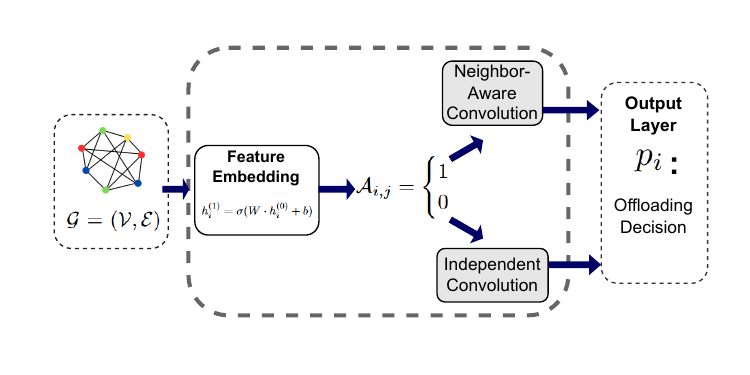}
  \caption{GNN-based Adaptive Decision Network}
  \label{fig:gnn_model}
\end{figure}

\section{Experiments and Evaluation}\label{section6}

\subsection{Dataset generation}\label{data_generation}

Our experiment utilized a dataset generated by solving the optimal solution from Section \ref{section4} using a standard optimization solver (Gurobi). The data set was created by running $5,000$ runs, resulting in a total of $1,375,000$ samples (task requests). 

To partition the dataset for model evaluation, we employed the $k$-fold cross-validation technique with $k=10$. This method divided the dataset $\mathbb{D}$ into 10 equally sized, non-overlapping subsets such that:

\begin{equation}
    \mathbb{D} = \bigcup_{i=1}^{10} \mathbb{D}_i, \quad \mathbb{D}_i \cap \mathbb{D}_j = \emptyset \ \text{for} \ i \neq j.
\end{equation}

For each training session $t$ ($t = 1, \ldots, 10$), one fold $\mathbb{D}_t$ was used as the validation set, while the remaining folds were combined to form the training set $\mathbb{D}_{\text{train}}^{(t)}$:

\begin{equation}
  \mathbb{D}_{\text{train}}^{(t)} = \bigcup_{\substack{i=1 \\ i \neq t}}^{10} \mathbb{D}_i, \quad \mathbb{D}_{\text{val}}^{(t)} = \mathbb{D}_t.  
\end{equation}

This resulted in 10 training sessions, each using a different subset for validation.

\subsection{Experiment Environment}

We utilized an SDN-based COIN-MEC network topology consisting of 12 COIN nodes supporting a single MEC, as depicted in Figure \ref{fig1:Experimental_Setup}. The upper layer comprises four COIN nodes interconnected in a fully meshed topology, each linked to the MEC. The COINS have different processing capabilities as reported in Table \ref{tab:parameter_settings}. WD's offload their tasks through set of APs to selected node(s).

The following settings obtain the Optimal solution: The simulation and model training are being run using Spyder (Integrated Development Environment) IDE \cite{raybaut2009spyder} for Python programming language, and the solver used is Gurobi 11.0.0 \cite{gurobi} and Pyomo framework \cite{hart2011pyomo,bynum2021pyomo}.

\subsection{Performance Metrics}

Table \ref{tab:metrics} presents the performance metrics of the models under evaluation. Our proposed GNN model is compared against previously employed supervised machine learning (ML) models from the literature \cite{lia2022}, specifically the Multi-Layer Perceptron (MLP) and Decision Tree (DT). The results demonstrate that the GNN model consistently outperforms these conventional models across all evaluated metrics.

In terms of accuracy, defined as the ratio of correctly predicted instances to the total number of observations, the GNN achieved an  accuracy of 92.44\%. This represents a significant improvement over the MLP and DT models, which achieved accuracies of 76.25\% and 70.16\%, respectively. 

\begin{table}[h]
  \centering
  \caption{Comparison of Metrics for GNN, MLP, and DT}
  \label{tab:metrics}
  \begin{tabular}{lccc}
    \hline
    \textbf{Metrics} & \textbf{GNN} & \textbf{MLP} & \textbf{DT} \\
    \hline
    Accuracy & \textbf{0.9244} & 0.7625 & 0.7016 \\
    Macro Average Precision & \textbf{0.9207} & 0.7122 & 0.7297 \\
    Macro Average Recall & \textbf{0.9244} & 0.7625 & 0.7016 \\
    Macro Average F1-score & \textbf{0.9196} & 0.7249 & 0.6731 \\
    \hline
  \end{tabular}
\end{table}

\subsection{Time Complexity analysis}

\begin{figure}[h]  % 'h' option tells LaTeX to try to place the figure "here"
  \centering
  \includegraphics[width=1.0\linewidth]{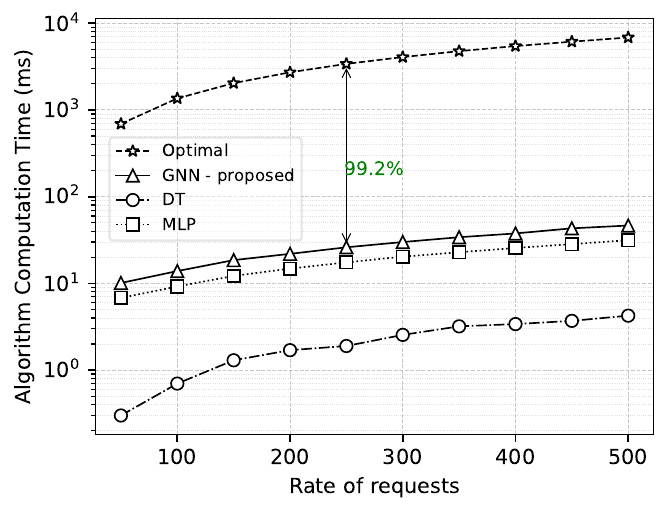}  % Adjust the width as needed
  \caption{Time complexity analysis of Algorithm computation of the ML models as compared to the optimal solution}
  \label{fig:figure4}
\end{figure}

Figure \ref{fig:figure4} analyze the time complexity of the proposed model against the solution obtained from standard optimization solver under heavy computation demand against the rate of requests. The time complexity of algorithm computation for the proposed GNN model compared to the optimal solution demonstrates that the optimal solution computes approximately 100 times slower than the GNN model. Moreover, this difference substantially increases as the number of rendering requests escalates. This is because of the exhaustive search to achieve the optimal solution. 

The simulations were conducted on a system equipped with a 12th Generation Intel(R) Core(TM) i5-12400F processor operating at 2.50 GHz, 16 GB of RAM, and a 500 GB hard drive.

\subsection{System Performance}

We evaluate the system performance from the perspective of the SDN controller by defining the system performance gain (PG) as the ratio of the system cost obtained using the optimization solver to the system cost achieved by the proposed GNN model:

\begin{equation}
    PG = \frac{C_{\text{GNN}}}{C_{\text{solver}}}
\end{equation}

In this study, we analyze the system performance gain (PG) for the proposed GNN model, a MEC-only system, and the compared supervised ML models.

\begin{figure}[h]  % 'h' option tells LaTeX to try to place the figure "here"
  \centering
  \includegraphics[width=1.0\linewidth]{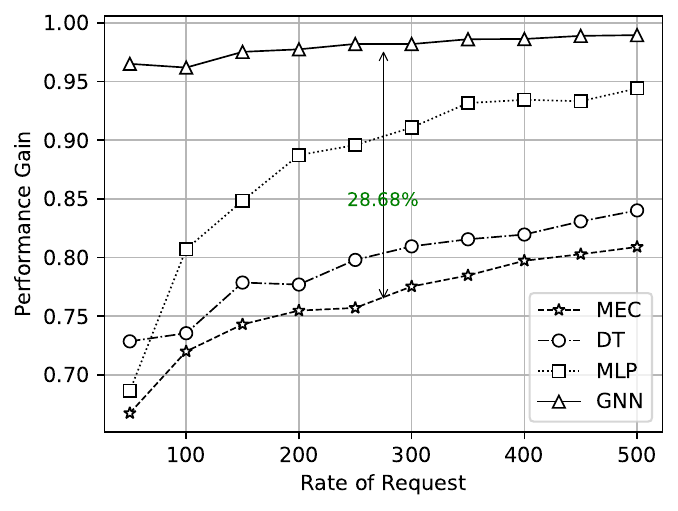}  % Adjust the width as needed
  \caption{Performance Gain vs Rate of Request}
  \label{fig:figure5}
\end{figure}

We first show PG as a function of the Rate of Request in figure \ref{fig:figure5}. We observe that the GNN achieves better performance gain than the other models, This is attributed to the GNN's ability to predict outcomes that closely approximate the solution obtained through the optimization solver. This effect is especially evident when there are more WD's sending requests, as a higher volume of requests provides the GNN with more data to learn from, thereby enhancing its prediction accuracy. Finally, we observe that the MEC-only system has lower performance gain, this is due to the amount of data exchanged in the network domain, because all input data need to traverse the network in order to reach the MEC for execution.

\subsection{TASK SPLITTING AND NON-SPLITTING SCENARIOS}

Different tasks have varying tolerable delay thresholds based on their requirements. To model these diverse requirements, we analyze system performance under two distinct $D_{\max}^k$ configurations with ranges [10, 150]ms and [50, 150]ms, with bound values aligned with those commonly used in literature \cite{wang2019delay, tocze2019orch}. 

\begin{figure}[h]  % 'h' option tells LaTeX to try to place the figure "here"
  \centering
  \includegraphics[width=1.0\linewidth]{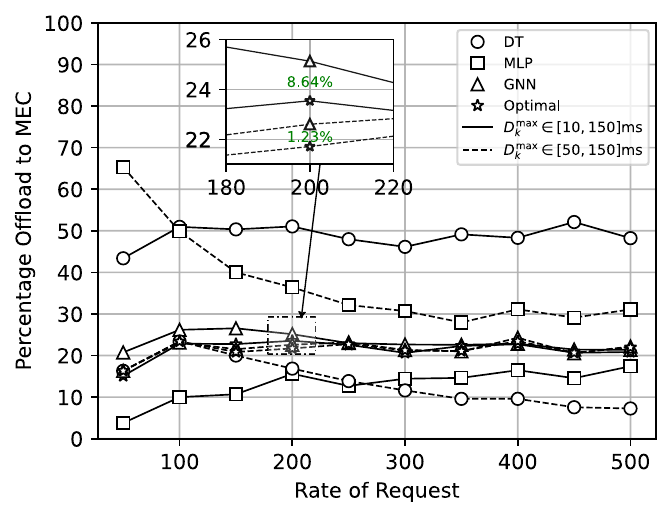}  % Adjust the width as needed
  \caption{Percentage offload to MEC with task splitting}
  \label{fig:figure6}
\end{figure}

Figure \ref{fig:figure6} illustrates the percentage of tasks offloaded to the MEC under the task-splitting scenario, where tasks are distributed across two COIN nodes. The results indicate that approximately 20\% of tasks are offloaded to the MEC across all rates of requests. This limited offloading is attributed to the ability of tasks with low tolerable delay to utilize the computational resources of the two COIN nodes effectively, thereby reducing the necessity for offloading to the MEC.

\begin{figure}[h]  % 'h' option tells LaTeX to try to place the figure "here"
  \centering
  \includegraphics[width=1.0\linewidth]{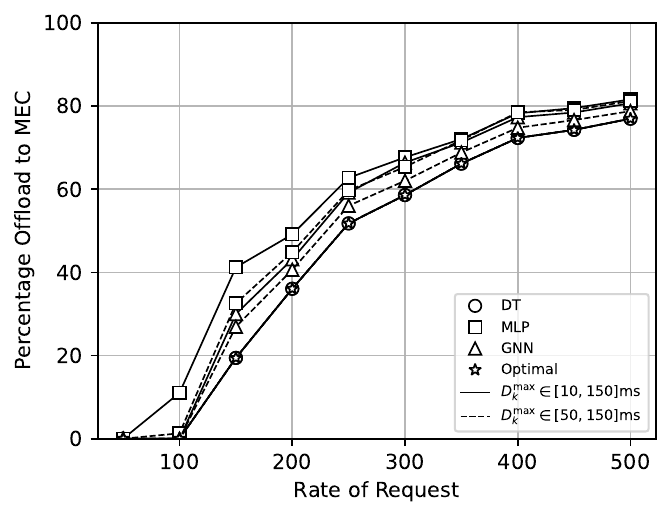}  % Adjust the width as needed
  \caption{Percentage offload to MEC with no task splitting}
  \label{fig:figure7}
\end{figure}

Figure \ref{fig:figure7}, on the other hand, illustrates the percentage of tasks offloaded to the MEC under the no-splitting scenario, where each task is assigned to a single COIN node without distribution. The results reveal a significantly higher percentage of tasks being offloaded to the MEC as number of requests increases. This increase is primarily due to the limited computational resources available at a single COIN node, which necessitates offloading tasks, particularly those with stricter delay constraints, to the MEC to meet their tolerable delay thresholds.

\begin{table}[ht]
    \centering
    \caption{Hyperparameter settings for training of GNN}
    \begin{tabular}{ll}
        \hline
        \textbf{Setting} & \textbf{Value} \\
        \hline
        Model type & Graph Convolutional Network (GCN)\\
        Number of features & 27 \\
        Activation function & ReLU \\
        Layers & 2 \\
        Optimizer & Adam \\
        Epochs & 200 \\
        Learning rate & 0.01 \\
        \hline
    \end{tabular}
\end{table}

\subsection{Computational Cost}

\begin{figure}[h]  % 'h' option tells LaTeX to try to place the figure "here"
  \centering
  \includegraphics[width=1.0\linewidth]{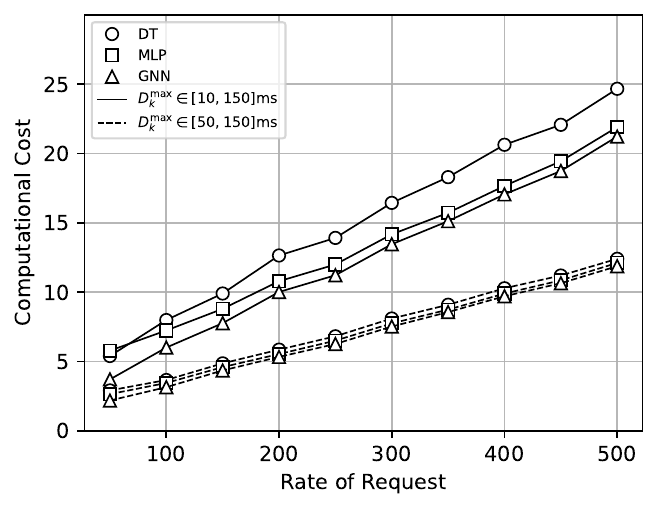}  % Adjust the width as needed
  \caption{Computational Complexity}
  \label{fig:figure8}
\end{figure}

Figure \ref{fig:figure8} shows the total network usage $C^n$ as a function of Request Rate for the GNN model against DT and MLP under $D_{\max}^k$ of range [10, 150] and [50, 150].

Due to the amount of data exchanged within the network, the network costs increases in each case as more users are sending requests. We can also note that when the strictness of the delay is relaxed, the amount of network resources used is lower, most of the tasks can be executed by the COINs which means that tasks are processed closer to the source, minimizing data transfer across the network.

\section{Conclusion and Discussion}\label{section7}
Considering the heavy training demands, there will be limited capabilities for the COINS during task computation. Since the task allocation problem is an online problem, i.e tasks arrive continuously and decisions must be made quickly. Therefore, we employed GNN to solve this work's formulated task placement problem offline. Eventhough this technique require a lot of training data and take time to train, this training can be done beforehand. The actual process of deciding where to send a task is very fast once the system is trained.

We conclude with discussing two potential extensions of our work. First, we discuss the applicability of our model to metaverse divisible tasks \cite{10366259}. Our model captures well the queuing and execution delay of these tasks at both the COIN nodes and the MEC. However transmission time between Wireless Devices (WDs), Access Points (APs), and COIN nodes were not considered. We leave this as an interesting avenue for future work.

Second, we discuss offloading tasks to the COINs to support the MEC. This approach is particularly beneficial in scenarios with high task arrival rates or limited MEC resources. However, this work does not consider the possibility of computing tasks locally on the Wireless Devices (WDs). 

Our work can also be applicable in other domains with delay-constrained requiring high performance computing. This includes online gaming, holographic, virtual reality (VR), and augmented reality (AR) applications, which similarly demand efficient and real-time processing to ensure immersive user experiences. Future studies should enhance the model’s scalability and robustness, vital for its effectiveness in extensive, complex metaverse settings.

\bibliographystyle{ieeetr}
\bibliography{main}

\end{document}